\documentclass[a4paper,10pt,twoside,dvips]{article}
\usepackage{amssymb, latexsym, amsmath, epsfig, subfigure}
\begin{document}

\title{Singular shell embedded into a cosmological model}

\author{\O yvind Gr\o n\thanks{e-mail: Oyvind.Gron@iu.hio.no}\\
Oslo College, Faculty of Engineering and\\
Dep. of Physics, University of Oslo
\and
Peter D. Rippis\thanks{e-mail: p.d.rippis@fys.uio.no}\\
Dep. of Physics, University of Oslo}
\date{\today}
\maketitle

\begin{abstract}
We generalize Israel's formalism  to cover singular shells
embedded in a non-vacuum Universe. That is, we deduce the relativistic
 equation of motion for a thin shell embedded in a Schwarzschild/Friedmann-Lema$\Hat{\mbox{\i}}$tre-Robertson-Walker spacetime.
Also, we review the embedding of a Schwarszchild mass into a cosmological model
using ``curvature'' coordinates 
and  give solutions with (Sch/FLRW) and without the embedded mass (FLRW). 
\end{abstract}

\section{Introduction}
The first applications of Israel's relativistic theory of singular surfaces
\cite{is1, bais} were concerned with neutral and charged surfaces layers in static spacetimes with or without a cosmological constant \cite{be1} - \cite{ku}.
Surface layers with
different equations of state, i.e. different relationships between pressure and density were considered, for example domain walls \cite{guth}. Also,  methods to quantize a shell in general relativity has been developed and studied for static spacetimes \cite{cor, be}. Lately
Israel's theory has been applied to three-dimensional branes in a five-dimensional bulk \cite{lan,sms}.

In the present work we want to extend the application of Israel's formalism to cases with singular layers (two-dimensional branes) in expanding Universe models.
In the static case with a neutral surface layer there is Schwarzschild or Schwarzschild-de Sitter spacetime. For non-static spacetimes the relativistic
 equation of 
motion for surface layers has been given 
for Friedmann-Lema$\Hat{\mbox{\i}}$tre-Robertson-Walker (FLRW) spacetimes \cite{be2} - \cite{masa}.
In the more general case  we must consider spacetime outside a mass embedded 
in an expanding Universe model.
A description of this was developed by R. Gautreau \cite{gau2, wils}.

We shall therefore apply Israel's theory to a singular layer with the Gautreau metric outside the shell.
The equation of motion for the general case is deduced here.

A similar problem, with a slowly rotating shell in an almost FLRW dust universe model was considerd by Klein \cite{klein} without using  Israel's formalism and later by T. Dole$\check{\mbox{z}}$el, J. Bi$\check{\mbox{c}}$$\acute{\mbox{a}}$k 
and N. Deruelle
\cite{dbd} using  Israel's formalism. They restricted their treatments by considering only a shell comoving with the  cosmic dust outside the shell, and with  Minkowski spacetime inside the shell. In our treatment we consider a non-rotating shell that  needs not be comoving with the cosmic fluid.     

\section{Schwarzschild mass embedded into a cosmological model}\label{sec:imbedd}

Embedding  a Schwarzschild mass, $m$,  into a cosmological model 
is most easily done in
``curvature'' coordinates. That is, in coordinates for 
which the radial coordinate $R$ gives an angular part $R^{2}d\Omega^{2}$.
The embedding of a Schwarzschild mass into  a spatially flat FLRW 
Universe model was given by R. Gautreau in \cite{gau2}. Also, a thorough
investigation of  the FLRW models with vanishing cosmological constant,
$\Lambda=0$, in these coordinates has been performed by 
 R. Gautreau, see \cite{gau3} and references therein.
In this section we give a short review. Then we give the equations for  embedding a 
Schwarzschild mass into a general FLRW Universe model and provide
solutions with and without the embedded mass.  

\subsection{The metric}
For a spherically symmetric spacetime we can write a  general metric
as:

\begin{equation}\label{eq:diagmetTR}
ds^{2}=-A(R,T)f^{2}(R,T)dT^{2}+ \frac{dR^{2}}{A(R,T)}+R^{2}d\Omega^{2},
\end{equation}
where
\begin{equation}
d\Omega^{2}=d\theta^{2}+\sin^{2}{\theta}d\phi^{2}.
\end{equation}
Here $A$ and $f$ are functions of $T$ and $R$ that are settled 
by Einstein's field
equations. 
The physical interpretation of the time coordinate $T$ is that it measures the
time on clocks that are located at points for which $R=constant$
relative to our chosen origin $R=0$.\footnote{The time laps in 
T and the time laps recorded on clocks
that measures proper time $\tau$ ($ds^{2}=-d\tau^{2}$)
for $R,\theta,\phi=constant$ are related by
$dT = \frac{1}{f\sqrt{A}}d\tau$.} 
In the FLRW Universe models one can define a global time coordinate.
Hence, we wish to record time on a geodesically moving clock. 
To transform to a geodesically moving clock we consider the geodesic equation,
$V^{\mu}_{~;\nu}V^{\nu}=0$, 
for a radially moving clock, i.e $d\theta=d\phi=0$; where  $V^{\mu}$ is the tangent vector to the geodesic curve. For timelike 
geodesics, $V^{\mu}V_{\mu}=-1$,\footnote{We use units where the speed of light
and the gravitational constant are set to unity, $c=G=1$.} 
the solution is found to be 
(see \cite{gau2}):

\begin{eqnarray}\label{eq:radgeoT}
V^{\mu}(T,R)\equiv \frac{dx^{\mu}}{d\tau}
=\left(\frac{\mathcal{E }}{fA},\:\varsigma(\mathcal{E}^{2}-A)^{\frac{1}{2}},\:0,\:0\right)\;,\;
\; \varsigma=\pm 1,
\end{eqnarray}
with the conditions (coordinate transformations):

\begin{eqnarray}\label{eq:tranradgeot1}
t_{,\mbox{\tiny{R}}}&=&- \varsigma A^{-1}(\mathcal E^{2}-A)^{\frac{1}{2}}\\
t_{,\mbox{\tiny{T}}}&=&\mathcal E f, \label{eq:tranradgeot2}
\end{eqnarray}
where $\mathcal E$ is an energy parameter for the reference particles, i.e. the geodesic clocks. $\mathcal E$ depends on our choice of reference system, and can be used to describe the open, closed and flat FLRW Universe models.
The time coordinate $t$ measures the time recorded on clocks moving 
on radial geodesics, i.e. $t$ 
measures  proper time, $t=\tau$, and thus on radial geodesics we have
$ds^{2}=-dt^{2}$.
The sign given by $\varsigma$ indicates whether the geodesic clock is 
moving with increasing R ($ \varsigma= + 1$) or 
with decreasing R ($ \varsigma=- 1$).
So that for  an expanding Universe we have $ \varsigma=+1$.
Now we can  make a coordinate transformation from $(T,R)$ to $(t,R)$ 
coordinates. In $(t,R)$ coordinates the radial geodesics are described by
the 4-velocity
\begin{eqnarray}\label{eq:radgeot}
V^{\mu}(t,R)&=&(1,\: \varsigma (\mathcal E^{2}-A)^{\frac{1}{2}},\:0,\:0).
\end{eqnarray}
The resulting form of the 
line element is:
\begin{eqnarray}\label{eq:metradgeot}
ds^{2}&=&\mathcal E^{-2}\Big(-Adt^{2}+ dR^{2}-2 \varsigma(\mathcal E^{2}-A)^{\frac{1}{2}}
dRdt\Big)+ R^{2}d\Omega^{2} \nonumber\\
     &=&-d t^{2}+\mathcal E^{-2}\Big(dR- \varsigma(\mathcal E^{2}-A)^{\frac{1}{2}}
     dt\Big)^{2}+ R^{2}d\Omega^{2}.
\end{eqnarray}
Because $t$ measures time on clocks moving relative to $R=constant$ 
the metric is  non-diagonal.
The cosmic particles, e.g. the galaxies are 
assumed to follow  radial geodesics. 
Without the embedded Schwarzschild mass, i.e.
 $m=0$, we require the Universe to be isotropic and homogeneous. The choice of coordinates ensures the isotropic condition. For a flat Universe model the 
homogeneity condition settles $\mathcal E$ \cite{gau2}. First we note that for
$m=0$ the coordinates are required to reduce to the Minkowskian form at 
$R=0$, i.e. $A=f=1$. Then from  (\ref{eq:radgeoT}) we find that at $R=0$
 the energy parameter $\mathcal E$ is given by the Lorentz factor $\gamma$:
$\mathcal E=(1-v_{_{0}}^{2})^{-\frac{1}{2}}$, where $v_{0}$ is the coordinate velocity, $\frac{dR}{dT}$, of the reference particles at $R=0$; and because the space-time metric here is flat this is the velocity measured by an observer at 
$R=0$. Hence, $\mathcal E$ gives the energy to rest mass ratio for the cosmic
particles. Particles with $\mathcal E<1$ will never reach $R=0$, thus making a hole in the cosmic fluid.
While for
 $\mathcal E>1$ they will have some velocity, $v_{_{0}}^{2}>0$, at $R=0$.
This makes $R=0$ a source ($\varsigma=+1$) or a sink ($\varsigma=-1$) for cosmic particles. For a homogeneous Universe $R=0$ cannot have any special
 significance. Hence, we are lead to

\begin{equation}
\mathcal E=1
\end{equation}
for a spatially flat Universe. If we embed a 
Schwarzschild mass into this model we get an inhomogeneous  model, but we still have $\mathcal E=1$ \cite{gau2}. Gautreau argues that 
for a vacuum Universe, i.e. the de Sitter model to have a physical reference system we cannot set $\mathcal E=1$. The vacuum Universe in the non-diagonal metric (\ref{eq:metradgeot}) is discussed in  \cite{gau1}. For a thorough investigation of the kinematics of the de-Sitter Universe  see \cite{ergr95}.

For a flat Universe model, i.e. $\mathcal E=1$, 
the surface of simultaneity $dt=0$ in (\ref{eq:metradgeot})
gives the flat metric.

\begin{equation}
dl^{2}=dR^{2}+R^{2}d\Omega^{2}.
\end{equation} 
In this case $R$ measures the actual distance between the cosmic particles.

We shall now  consider 
$\mathcal E$ for open and closed models. 
To get a feeling for the coordinates in  (\ref{eq:metradgeot})
we relate them to the commonly used comoving coordinates 
in cosmology. In these coordinates the metric has the form

\begin{equation}\label{eq:stdcos}
ds^{2}=-dt^{2}+a(t)^{2}\bigg(\frac{dr^{2}}{1-kr^{2}}+r^{2}d\Omega^{2}\bigg)
\end{equation}
where $a(t)$ is the expansion factor and the radial coordinate $r$ is 
constant along the trajectory of the cosmic particles, e.g. galaxies, and $t$ is the time measured on clocks moving with them. The sign $k$ gives the spatial topology (i.e. the global structure of the $t=constant$ surface):
$k=+1$ leads to a spherical (closed) geometry, $k=-1$ gives a hyperbolic (open) geometry and
$k=0$ describes an Euclidean (flat) geometry.

The time coordinates in  (\ref{eq:metradgeot}) and  (\ref{eq:stdcos})
measure time on clocks moving with the cosmic particles. Thus   
we identify the time coordinates in (\ref{eq:metradgeot}) and 
(\ref{eq:stdcos}). The radial coordinates are related by $R=ar$ and 
the energy parameter, $\mathcal E$, for the reference particles is given by
$\mathcal E^{2}=1-kr^{2}=1-kC(t,R)$, where $C(t,R)$ is a function that is constant along the radial geodesics, i.e. $\mathcal E$  
must be constant along the geodesics. 
The Einstein equations  give this function up to a constant factor. 
The Hubble factor, 
$H\equiv \frac{1}{a}\frac{da}{dt}$, is now given by

\begin{equation}
\frac{dR}{dt}=HR.
\end{equation}

\subsection{The Einstein equations for curvature coordinates}

The Einstein equations with cosmological constant 
$G^{\mu}_{~\nu}+\Lambda\delta^{\mu}_{~\nu}=8\pi T^{\mu}_{~\nu}$ which 
 we will need to deal with are:

\begin{eqnarray} 
&&\frac{\partial (R(1-A))}{\partial R}=
-8\pi R^{2} T^{t}_{~t} +\Lambda R^{2} \label{eq:imbeina}\\ 
                                            \nonumber \\
&&\frac{\partial (R(1-A))}{\partial t}=8\pi R^{2} T^{R}_{~t} 
\label{eq:imbeinb} \\ 
 \nonumber \\
&&\frac{\partial (R(1-A))}{\partial R}+
 \varsigma(\mathcal E^{2}-A)^{-\frac{1}{2}}\frac{\partial (R(1-A))}{\partial t} + 
                                            \label{eq:imbeinc}   \\
&&2\varsigma\left(\frac{RA}{\mathcal E}\right)(\mathcal E^{2}-A)^{-\frac{1}{2}}
\bigg(\frac{\partial \mathcal E}{\partial t}+
 \varsigma(\mathcal E^{2}-A)^{\frac{1}{2}}\frac{\partial \mathcal E}{\partial R}\bigg)=
-8\pi R^{2} T^{R}_{~R}+\Lambda R^{2}.  \nonumber
\end{eqnarray}
These equations are valid in coordinates $x^{\mu}=(t,R,\theta,\phi)$
for an arbitrary energy-momentum tensor $T^{\mu}_{~\nu}$. The field
equations involving second-order derivatives are the angular part, for which
$T^{\theta}_{~\theta}=T^{\phi}_{~\phi}$; these equations are contained 
in the divergence of the Einstein tensor, $G^{~\mu}_{\nu~;\mu}=0$.

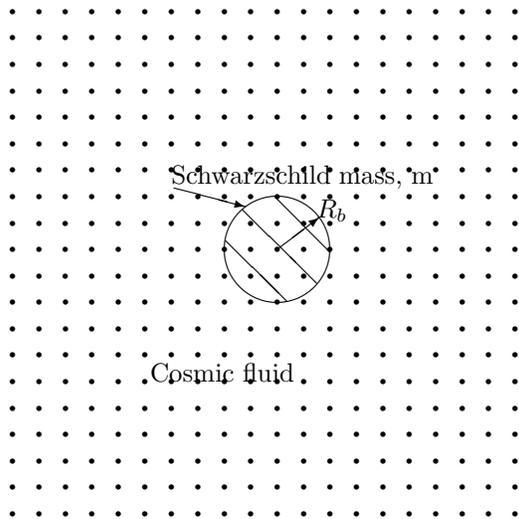
\begin{figure} 
\begin{center}
\begin{picture}(200,200)(0,0)
\put(100,100){\circle{50}}
\put(100,100){\vector(4,3){16}}
\put(115,112){$R_{b}$}
\put(61,123){\vector(4,-1){27}}
\put(60,125){Schwarzschild mass, m}
\put(87,115){\line(1,-1){28}}
\put(80.7,103.3){\line(1,-1){23}}
\put(100,119){\line(1,-1){20}}
\multiput(0,0)(10,10){20}{\circle*{2}}
\multiput(10,0)(10,10){19}{\circle*{2}}
\multiput(0,10)(10,10){19}{\circle*{2}}
\multiput(20,0)(10,10){18}{\circle*{2}}
\multiput(0,20)(10,10){18}{\circle*{2}}
\multiput(30,0)(10,10){17}{\circle*{2}}
\multiput(0,30)(10,10){17}{\circle*{2}}
\multiput(40,0)(10,10){16}{\circle*{2}}
\multiput(0,40)(10,10){16}{\circle*{2}}
\multiput(50,0)(10,10){15}{\circle*{2}}
\multiput(0,50)(10,10){15}{\circle*{2}}
\multiput(60,0)(10,10){14}{\circle*{2}}
\multiput(0,60)(10,10){14}{\circle*{2}}
\multiput(70,0)(10,10){13}{\circle*{2}}
\multiput(0,70)(10,10){13}{\circle*{2}}
\multiput(80,0)(10,10){12}{\circle*{2}}
\multiput(0,80)(10,10){12}{\circle*{2}}
\multiput(90,0)(10,10){11}{\circle*{2}}
\multiput(0,90)(10,10){11}{\circle*{2}}
\multiput(100,0)(10,10){10}{\circle*{2}}
\multiput(0,100)(10,10){10}{\circle*{2}}
\multiput(110,0)(10,10){9}{\circle*{2}}
\multiput(0,110)(10,10){9}{\circle*{2}}
\multiput(120,0)(10,10){8}{\circle*{2}}
\multiput(0,120)(10,10){8}{\circle*{2}}
\multiput(130,0)(10,10){7}{\circle*{2}}
\multiput(0,130)(10,10){7}{\circle*{2}}
\multiput(140,0)(10,10){6}{\circle*{2}}
\multiput(0,140)(10,10){6}{\circle*{2}}
\multiput(150,0)(10,10){5}{\circle*{2}}
\multiput(0,150)(10,10){5}{\circle*{2}}
\multiput(160,0)(10,10){4}{\circle*{2}}
\multiput(0,160)(10,10){4}{\circle*{2}}
\multiput(170,0)(10,10){3}{\circle*{2}}
\multiput(0,170)(10,10){3}{\circle*{2}}
\multiput(180,0)(10,10){2}{\circle*{2}}
\multiput(0,180)(10,10){2}{\circle*{2}}
\multiput(190,0)(10,10){1}{\circle*{2}}
\multiput(0,190)(10,10){1}{\circle*{2}}
\put(52,50){Cosmic fluid}
\end{picture}
\caption{Schwarzschild mass embedded in a cosmic fluid.}\label{fig:imbedd}
\end{center}
\end{figure}

To proceed  an energy-momentum
tensor for the Universe with an embedded Schwarzschild mass is assumed.
One may imagine the galaxies to be particles of a fluid and
that this cosmic fluid fills the whole spacetime.\footnote{The embedding of a Schwarzschild mass into a cosmological model has also been considered by 
Einstein and Straus \cite{es} (the Swiss Cheese model). They looked at a cosmic dust fluid with a vacuum
region ($\Lambda=0$) with a Schwarzschild mass at the center. And showed that one can join the Schwarzschild metric smoothly onto the cosmic metric
(\ref{eq:stdcos}). An explicit form of this metric was found by 
Sch$\ddot{\mbox{u}}$cking \cite{schu} using ``curvature'' coordinates.
For recent developments on this embedding see e.g. \cite{klein,bo,ma}.}
The embedded 
Schwarzschild mass, m, is placed with its center at $R=0$, and with boundary at
$R=R_{b}$, see figure (\ref{fig:imbedd}).
 As mentioned above the particles of the cosmic fluid are assumed to 
follow the radial geodesics $V^{\mu}$.
The cosmic fluid is described as an ideal fluid. 
Thus, outside $R_{b}$ the energy-momentum tensor is:

\begin{equation}\label{eq:cosflu}
T^{\mu}_{~\nu}=(\rho+p)V^{\mu}V_{\nu} +p\delta^{\mu}_{~\nu},\;\; R> R_{b},
\end{equation}
where $\rho$ is the mass density and $p$ the pressure of the cosmic fluid.
Inserting $V^{\mu}$ given in
(\ref{eq:radgeot}) into the energy-momentum tensor 
(\ref{eq:cosflu})  gives the components:  
 
\begin{eqnarray}
&&T^{t}_{~t}=-\rho, \quad T^{R}_{~R}=T^{\theta}_{~\theta}=T^{\phi}_{~\phi}=p, \nonumber\\ 
&&\label{eq:emcomp}\\
&&T^{t}_{~R}= 0, \quad T^{R}_{~t}=- \varsigma(\mathcal E^{2}-A)^{\frac{1}{2}}(\rho+p).\nonumber
\end{eqnarray}

Inside $R_{b}$ we will assume that the $tt$ component of the
energy-momentum tensor can be written as

\begin{equation}\label{eq:enmoin}
T^{t}_{~t} =-\rho+\rho^{t}_{~t}, \;\; R\leq R_{b},
\end{equation}
where $\rho$ is the energy density of the cosmic fluid and $\rho^{t}_{~t}$
is the part of the energy-momentum tensor
giving the energy density  of the embedded Schwarzschild 
mass. If we assume that  $\rho^{t}_{~t}$ is time independent this is the only component of the energy momentum tensor inside $R_{b}$ we will need.
  The constant embedded Schwarzschild mass, m, bounded by $R_{b}$ is 
defined by:

\begin{equation}\label{eq:defmsch}
m=-4\pi\int_{0}^{R_{b}} \rho^{t}_{~t} R^{2}dR.
\end{equation}
We note that setting $m=0$ leads to a
description of cosmology in ``curvature'' coordinates.
Let us also define a mass function $M(t,R)$ for the Universe:

\begin{equation}\label{eq:M}
M(t,R)=4\pi\int_{0}^{R} \rho(t, R) R^{2}dR,
\end{equation}
where the integration is over
a $t=constant$ surface.
In terms of $M$ the density is given by
\begin{equation}\label{eq:partM}
\frac{\partial M(t,R)}{\partial R}=4\pi R^{2} \rho(t, R),
\end{equation}
where $4\pi R^{2}$ gives the area of a sphere centered on $R=0$.

In the case where we have an embedded mass $m$ the density $\rho$ of the 
cosmic fluid will have 
some $R$ dependence and we cannot bring $\rho$ outside the integrals. 
But for  $m=0$ the cosmic fluid is homogeneous, and 
$\rho$ is only a function of $t$ and thus can be brought outside the 
integrals.
 
We now move on to solve the Einstein equations 
(\ref{eq:imbeina}-\ref{eq:imbeinc}) outside $R_{b}$, with an
energy-momentum tensor given by  (\ref{eq:emcomp}) and 
(\ref{eq:enmoin}). We start by integrating  (\ref{eq:imbeina}) over
a $t=constant$ surface. Using  (\ref{eq:defmsch}) and
 (\ref{eq:M}) this  gives the metric function in the form

\begin{equation}\label{eq:A}
A=1-\frac{2(m+M)}{R}-\frac{\Lambda}{3}R^{2},
 \end{equation}
with the metric given in (\ref{eq:metradgeot}). Note that in general $M$ is a function of $t$ and  R.

We need to determine M. Inserting (\ref{eq:A}) into (\ref{eq:imbeinb}) gives

\begin{eqnarray}\label{eq:fluxR}
T^{R}_{~t}&=&\frac{1}{4\pi R^{2}}\frac{\partial M}{\partial t}\\
&=&\frac{1}{R^{2}}\int_{0}^{R}\frac{\partial \rho}
{\partial t}R^{2}dR \nonumber
\end{eqnarray}
which gives the energy (mass) flux across a $R=constant$ surface. 
The $T^{R}_{t}$ component of the energy momentum tensor for an ideal fluid
 is  given in (\ref{eq:emcomp}). Equating these two expressions for 
$T^{R}_{t}$ and using  (\ref{eq:partM}) we find

\begin{equation}\label{eq:diffMvac}
\frac{\partial M}{\partial t}+
 \varsigma(\mathcal E^{2}-A)^{\frac{1}{2}}\frac{\partial M}{\partial R}=
-\varsigma 4\pi R^{2}p (\mathcal E^{2}-A)^{\frac{1}{2}},
\end{equation}
which is the partial differential equation determining $M$. From 
(\ref{eq:diffMvac}) and  (\ref{eq:partM}), assuming $p\geq -\rho$, we see that 
for an expanding Universe ($\varsigma=+1$) we have 
$\frac{\partial M}{\partial t}\leq 0$; and for a contracting Universe
($\varsigma=-1$) we get  $\frac{\partial M}{\partial t}\geq 0$.

To proceed from here we assume an equation of state $p=p(\rho)$ for 
the cosmic fluid. We assume that the cosmic fluid obeys the 
barotropic equation of state

\begin{equation}\label{eq:eos}
p=\omega\rho,\quad  \omega=\omega(t,R),
\end{equation}
e.g. $\omega=0$ gives a dust Universe,  $\omega=\frac{1}{3}$ describes a radiation dominated Universe and  $\omega=-1$ represents a vacuum Universe.
Using  (\ref{eq:partM})
we get
\begin{equation}\label{eq:diffMpwrho}
\frac{\partial M}{\partial t}+
\varsigma(1+\omega)
(\mathcal E^{2}-A)^{\frac{1}{2}}\frac{\partial M}{\partial R}=0.
\end{equation}  
Writing (\ref{eq:diffMvac})  in terms of $\rho$ \cite{gau2}
for a general Universe leads to  

\begin{equation}\label{eq:genuni}
R^{\frac{1}{2}} 
\int_{0}^{R} \frac{\partial \rho}{\partial \tau}R^{2}dR +\varsigma R^{2}(\rho +p)
\bigg(\mathcal E^{2} -1+8\pi 
\int_{0}^{R}  \rho R^{2}dR + 2m +\frac{\Lambda}{3}R^{3}\bigg)^{\frac{1}{2}}=0.
\end{equation}

Evaluating   (\ref{eq:diffMvac}) along the geodesics we find

\begin{equation}\label{eq:Mgeo}
\frac{d M}{d t}=
-4\pi R^{2} p \frac{d R}{d t}.
\end{equation} 
For dust, $p=0$, this gives $\frac{d M}{d t}=0$. So that
M is constant along the geodesics.

Consider the vacuum case. For Lorentz invariant vacuum energy the energy momentum tensor is proportional to the metric \cite{gron}, i.e. 
$T_{\mu\nu}=\rho_{_{\!\Lambda}}g_{\mu\nu}$ with the vacuum energy density 
$\rho_{_{\!\Lambda}}$ constant. Thus, this type of vacuum energy can be incorporated into the cosmological constant $\Lambda$; we have $\Lambda=8\pi\rho_{_{\!\Lambda}}$.
So that here the embedding of a Schwarszchild mass into the
 de Sitter Universe is given by $M=0$. The trivial solution of
(\ref{eq:diffMpwrho}) is $M=constant$, and if this is a global solution then from the definition of $M$
in  (\ref{eq:M}) and  (\ref{eq:partM}) this constant must be zero.    
Let now $\rho_{_{\!\Lambda}}$ be part of the energy momentum tensor, i.e. 
included in M. For vacuum,  $\omega=-1$, the second term in (\ref{eq:diffMpwrho}) or (\ref{eq:genuni}) vanish. This gives $\frac{\partial M}{\partial t}=0$
showing that M is time independent. For a homogeneous space $\rho_{_{\!\Lambda}}$   must then be constant.  
Let us also briefly look at the Einstein equations using the diagonal metric
  (\ref{eq:diagmetTR}). The above  equations become for an arbitrary $T^{\mu}_{~\nu}$: 

\begin{eqnarray}\label{eq:solAT}
&&A(R,T)=1 + \frac{8\pi}{R}\int_{0}^{R}T^{T}_{~T}  R^{2}dR -
\frac{\Lambda}{3}R^{2},\\
\label{eq:fluxRT}
&&T^{R}_{~T}=-\frac{1}{R^{2}}\int_{0}^{R}\frac{\partial T^{T}_{~T}}
{\partial T}R^{2}dR,\\
\label{eq:solf}
&&lnf^{2}=8\pi\int_{0}^{R}\frac{R}{A}(T^{R}_{~R}-T^{T}_{~T})dR.
\end{eqnarray}
The integration is now over a $T=constant$ surface. The coordinate transformation between the coordinate systems is given by 
(\ref{eq:tranradgeot1}) and (\ref{eq:tranradgeot2}). 
From  (\ref{eq:solf}) we see that all vacuum space-times have $f=1$.
And for a Schwarzschild mass embedded in a de-Sitter Universe the metric
(\ref{eq:diagmetTR}) is given by $A=1-\frac{2m}{R}-\frac{\Lambda}{3}R^{2}$.

Let us now consider $\mathcal E$. From the Einstein equation (\ref{eq:imbeinc}) and demanding that  
(\ref{eq:diffMvac}) is satisfied we find 

\begin{equation}\label{eq:diffkap}
\frac{\partial \mathcal E}{\partial t}+
 \varsigma(\mathcal E^{2}-A)^{\frac{1}{2}}\frac{\partial \mathcal E}{\partial R}=0.
\end{equation}
For geodesics this gives  $\frac{d\mathcal E}{dt}=0$, 
which shows that the Einstein equations require that $\mathcal E$ is a constant
along the streamlines of the cosmic fluid outside $R_{b}$. For a flat Universe model, with and without the embedded Schwarzschild mass,
this constant must be set to  $\mathcal E=1$. 
For a  Universe with a non zero spatial curvature we see from eq. 
(\ref{eq:diffMpwrho}) that $\mathcal E$ 
can be written in terms of $M$ for a dust Universe. 
To recover standard cosmology we require that $\mathcal E^{2}-1$ is proportional to 
$M$. Hence, we can write  

\begin{equation}\label{eq:Etop}
\mathcal E^{2}=1-k\frac{2M}{R_{i}},
\end{equation}
where $R_{i}=constant$. This constant depends on which galaxy we wish to follow and thus it 
determines M. E.g. for a closed ($k=+1$) Universe $R_{i}$ represents the maximum radius reached by a given galaxy. For further discussion of this and the open case see \cite{gau5}.

Next we will  consider the velocity and the acceleration of the cosmic
particles. From the geodesics in (\ref{eq:radgeot})
the coordinate velocity is in general given by 

\begin{equation}\label{eq:velgenR} 
\frac{dR}{dt}=(\mathcal E^{2}-A)^{\frac{1}{2}},
\end{equation}
with the Hubble parameter given by $H=\frac{1}{R}\frac{dR}{dt}$.
Inserting for A and $\mathcal E$ gives

\begin{equation}\label{eq:veldvacR} 
\bigg(\frac{dR}{dt}\bigg)^{2}=-k\frac{2M}{R_{i}}+\frac{2(m+M)}{R}+ 
\frac{\Lambda}{3}R^{2},
\end{equation}
where $M$ is constant  for a dust ($p=0$) Universe. 
Differentiating (\ref{eq:veldvacR}) and inserting  (\ref{eq:Mgeo})
with $p=0$ we get the acceleration for a dust Universe,

\begin{equation}\label{eq:accgenR} 
\frac{d^{2}R}{dt^{2}}=-\frac{m+M}{R^{2}}+\frac{\Lambda}{3}R.
\end{equation}
For $m=0$ we have a homogeneous cosmic fluid, thus we can find $M$ from considering the geodesics
(\ref{eq:veldvacR}) and (\ref{eq:accgenR}). 
If  $m\neq 0$ the cosmic fluid is inhomogeneous and we must find $M$ by 
treating the coordinates, $t$ and $R$, as independent variables, 
i.e. from eq. (\ref{eq:diffMvac}). 
 Also, we note that
 for  $m=0$ equations  (\ref{eq:veldvacR}) and (\ref{eq:accgenR}) 
are equivalent to the Friedmann equations, 
i.e. the dynamical equations obtained from Einsten's field equations 
when using the comoving coordinates in  (\ref{eq:stdcos}). 
For $\Lambda=0$ (\ref{eq:accgenR}) is identical to Newton's 
gravitational law, see \cite{gau5}.

For light the trajectories are given by $ds^{2}=0$, thus the paths of  radially moving light are described by

\begin{equation}\label{eq:lighttra} 
\frac{dR}{dt}=(\mathcal E^{2}-A)^{\frac{1}{2}}\pm \mathcal E.  
\end{equation}
For a closed Universe model, $k=+1$, in the limit 
$R_{i}\rightarrow 2M$ we have $\mathcal E^{2}\rightarrow 0$ giving a 
 singularity in the metric (\ref{eq:metradgeot}). In this case (\ref{eq:veldvacR}) approaches (\ref{eq:lighttra}), i.e. the galaxies approach the speed of light. This is the maximum size of the Universe\cite{gau5}.

\section{Solutions}
\subsection{$m=0$}
For a general Universe without the embedded Schwarzschild mass we can bring
$\rho$ outside the integrals above. 
The mass function, $M$, for  the Universe defined in  (\ref{eq:M})
becomes
\begin{equation}\label{eq:Mmnull}
M=\frac{4\pi}{3}R^{3}\rho(t).
\end{equation} 
For a dust Universe this gives the mass of the Universe inside R at time $t$.
In the following we shall consider a flat Universe model, $\mathcal E =1$, 
such that R measures the proper distance between the cosmic particles.  
Eq. (\ref{eq:genuni}) then reduces to:

\begin{equation}\label{eq:genU}
\frac{d\rho}{d t}=-\varsigma (\rho+p)(24\pi\rho+3\Lambda)^{\frac{1}{2}}.
\end{equation}  
Along the geodesics (see (\ref{eq:Mgeo})) eq. (\ref{eq:genU}) reads 

\begin{equation}\label{eq:veldrho}
\frac{d\rho}{dt}=-3(\rho+p)\frac{1}{R}\frac{dR}{dt},
\end{equation}  
independent of $\Lambda$ (and $\mathcal E$).
From  (\ref{eq:genU}) and  (\ref{eq:veldrho}) we  find a 
1-parameter family of solutions. We assume an equation of state
of the form $p=\omega\rho$ with $\omega=constant$. 
Integration of  (\ref{eq:veldrho}) then gives $\rho(t)$ in terms of the geodesics, $R(t)$:

\begin{equation}\label{eq:solveldrho}
\rho(t)=bR(t)^{-3(1+\omega)}.
\end{equation}
This expression for the density is in fact valid for all the FLRW Universe models, i.e. regardless of $\Lambda$ and $\mathcal E$.
For dust $\omega=0$ which inserted into  (\ref{eq:solveldrho}) gives
$\rho\propto R^{-3}$, i.e. M is constant, for radiation we have $\omega=\frac{1}{3}$ and
we get $\rho\propto R^{-4}$. 
For vacuum, $\omega=-1$,  we get $\rho=constant$  which is also obtained from
eq. (\ref{eq:genU}); the expansion of the vacuum Universe is found from 
eq. (\ref{eq:veldvacR}): 

\begin{alignat}{2}
R&\propto \cosh{\sqrt{\frac{\Lambda}{3}}t},&\qquad&\mathcal E < 1, \\
R&\propto e^{\sqrt{\frac{\Lambda}{3}}t},&\qquad&\mathcal E = 1,\\
R&\propto \sinh{\sqrt{\frac{\Lambda}{3}}t},&\qquad&\mathcal E > 1. 
\end{alignat}

In the following we assume that  $\omega\neq -1$.
We note that the density $\rho$ does not depend on the integration constant 
$b$. This constant depends on which geodesic (galaxy) we wish to follow, 
(see below). Thus, we can relate $b$ to the comoving coordinate $r$ in 
(\ref{eq:stdcos}). Normalizing the expansion factor, $a$, so that $a_{0}=1$, where the index $0$ refers to the present time, we get $b=r^{3(1+\omega)}\rho_{0}$. For a dust Universe M is constant and we find that $M=\frac{4\pi}{3}b$.
That is, $b$ is a measure of the mass inside a comoving radius $r=constant$. 
Alternatively, one can regard $b$ as a measure of the energy of a cosmic particle relative to $R=0$, $t=0$.
 
\begin{figure}[!]
\centering
\mbox{\subfigure[Einstein-de Sitter]{\epsfig{figure=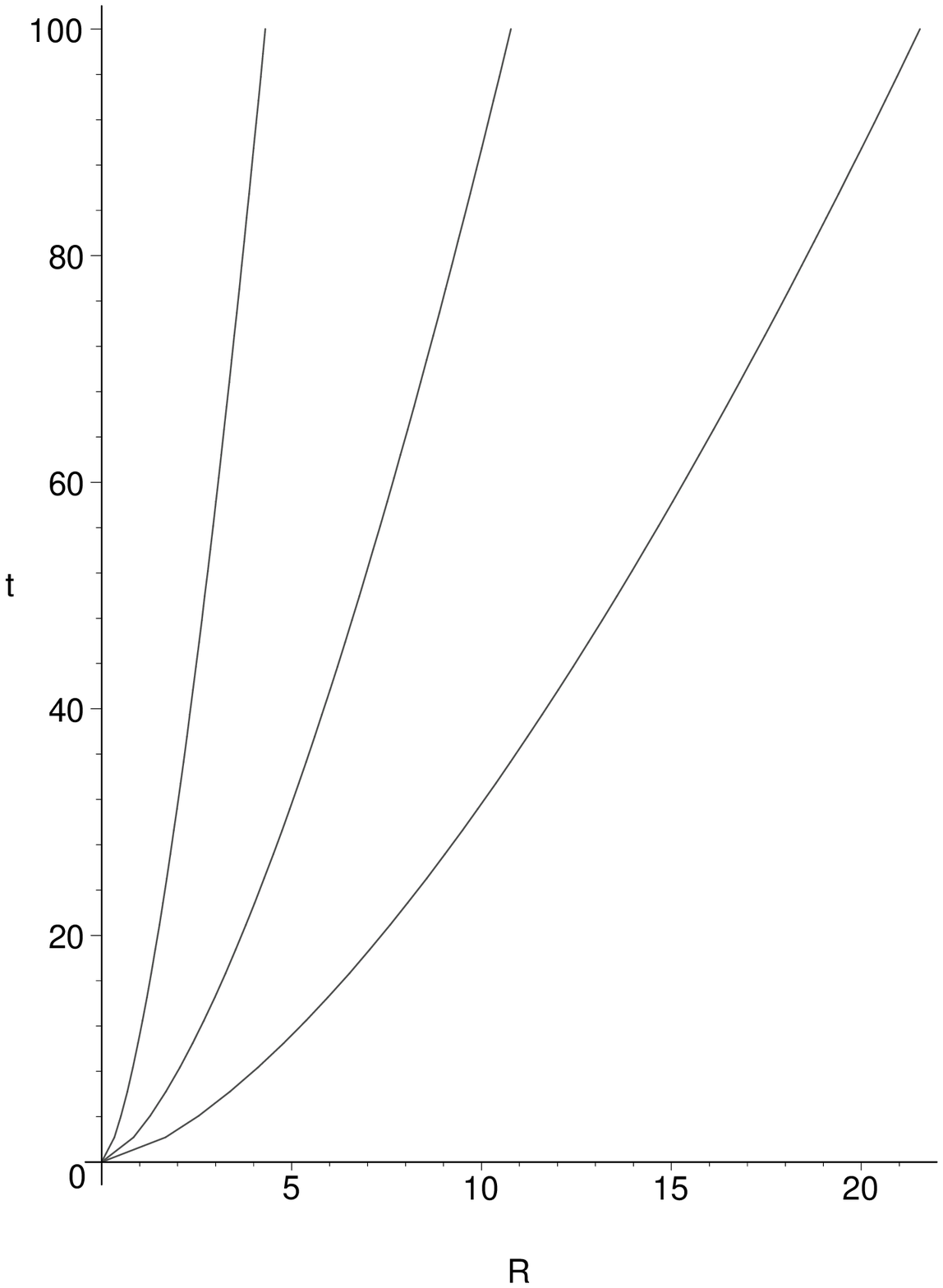,width=.45\textwidth}}
\hspace{2cm}
\subfigure[Friedmann-Lema$\Hat{\mbox{\i}}$tre]{\epsfig{figure=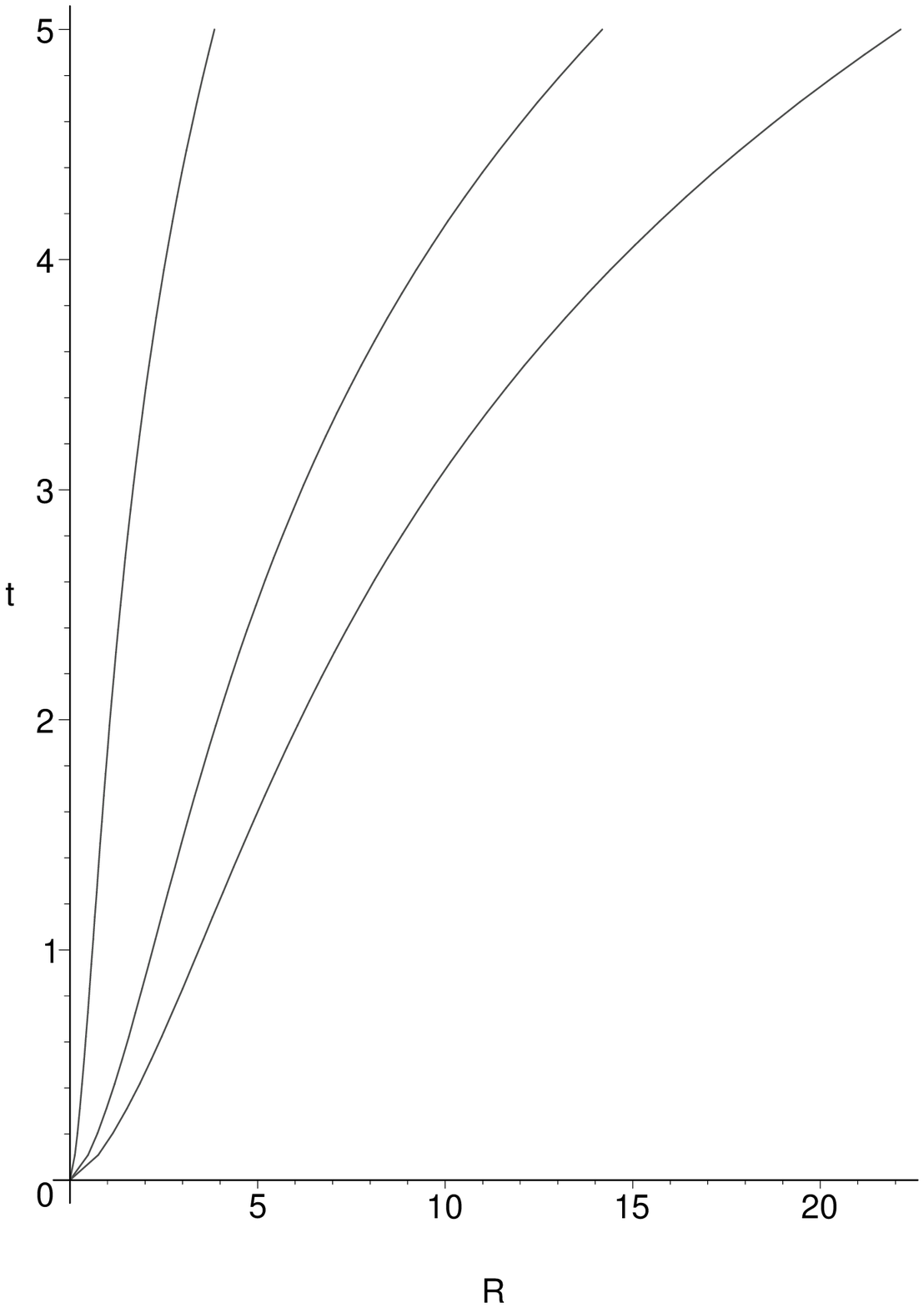,width=.45\textwidth}}}
\caption{Graphs showing the radial timelike geodesics for different values of $b$ in a Universe containing (a) dust, and (b) dust and vacuum energy.}\label{fig:geoem}
\end{figure}

Solving  eq. (\ref{eq:genU}) for $\Lambda=0$ gives
\begin{equation}\label{eq:rot}
\rho=\frac{1}{(1+\omega)^{2}6\pi t^{2}}.
\end{equation}
We have set the integration constant to zero so that the ¨big bang¨ is placed at  $R(t\!\!=\!\!0)\!=\!0$. 
Eq.  (\ref{eq:rot}) shows that the density of all ideal fluids, in a flat Universe, with $p=\omega\rho$ are proportional to
$t^{-2}$. 
Combining (\ref{eq:solveldrho}) and (\ref{eq:rot})
 we find the geodesics:

\begin{equation}\label{eq:soltra}
R=\Big(b(1+\omega)^{2}6\pi t^{2}\Big)^{\frac{1}{3(1+\omega)}}.
\end{equation}
The geodesics for for the Einstein-de Sitter model, $\omega=0$, 
are displayed in fig.\ref{fig:geoem}, where the geodesic $R=0$ represents the trajectory of e.g. our galaxy.
Inserting  (\ref{eq:rot}) and  (\ref{eq:Mmnull}) into  (\ref{eq:A}) leads to the metric:

\begin{equation}\label{eq:Aflat}
A=1-\left(\frac{2R}{3(1+\omega)t}\right)^{2}
\end{equation}
and the Hubble parameter
\begin{equation}\label{eq:Hflat}
H=(1-A)^{\frac{1}{2}}\frac{1}{R}=\frac{2}{3(1+\omega)t},
\end{equation}
where the line-element is given in (\ref{eq:metradgeot}).

The current standard model of the Universe is the flat 
Friedmann-Lema$\Hat{\mbox{\i}}$tre model, which is a Universe model with dust and Lorentz invariant vacuum energy; i.e. $\omega=0$ and $\Lambda>0$. A 
pedagogical presentation of this model is given in\cite{gr02}.

For a non zero $\Lambda$ there are two cases: $\Lambda> 0$ and $\Lambda<0$.
Integrating  (\ref{eq:genU}) for  $\Lambda> 0$ we get 

\begin{equation}\label{eq:rhonewstd}
\rho=\frac{\Lambda}{4\pi\cosh{\Big((1+\omega)\sqrt{3\Lambda}\:t\Big)}-4\pi}.
\end{equation}
Taking the series for $\cosh$ we get (\ref{eq:rot})   in the limit $\Lambda\rightarrow 0$.
Eq. (\ref{eq:rhonewstd}) can also be written  
\begin{equation} 
\rho=\frac{\Lambda}{8\pi \sinh^{2}{\Big((1+\omega)\frac{\sqrt{3\Lambda}}{2}
\:t}\Big)}. 
\end{equation}
Thus, from eq. (\ref{eq:A})  the metric (\ref{eq:metradgeot}) is given by
\begin{equation}\label{eq:ALflat}
A=1-\frac{\Lambda}{3} R^{2}\coth^{2}{\Big((1+\omega)\frac{\sqrt{3\Lambda}}{2}
\:t}\Big)
\end{equation}
and the Hubble parameter
\begin{equation}\label{eq:HFLflat}
H=\sqrt{\frac{\Lambda}{3}}\coth{\Big((1+\omega)\frac{\sqrt{3\Lambda}}{2}
\:t}\Big).
\end{equation}

From (\ref{eq:solveldrho}) the trajectories are

\begin{eqnarray} 
R&=&
\Bigg(b\frac{4\pi\cosh{\Big((1+\omega)\sqrt{3\Lambda}\:t\Big)}-4\pi}{\Lambda}
\Bigg)^{\frac{1}{3(1+\omega)}}\\ 
&=&\Bigg( b\frac{8\pi\sinh^{2}{\Big((1+\omega)\frac{\sqrt{3\Lambda}}{2}\:t}\Big)} 
{\Lambda}\Bigg)^{\frac{1}{3(1+\omega)}}
\end{eqnarray} 
which are shown for the Friedmann-Lema$\Hat{\mbox{\i}}$tre model, i.e. $\omega=0$,
 in fig.\ref{fig:geoem}.
Eq. (\ref{eq:accgenR}) then implies  that the expansion of the Universe becomes
accelerated for
\begin{equation}
\Lambda>4\pi\rho,
\vspace*{2mm}%
\end{equation}
or in terms of the Lorentz invariant vacuum energy density:
$\rho_{\!_{\Lambda}}>\frac{1}{2}\rho$, \cite{gr02}.

Solving (\ref{eq:genU}) for $\Lambda<0$ we  obtain the solution

\begin{equation}
\rho=\frac{-3\Lambda}{24\pi\cos^{^{2}}{\Big(-\frac{\sqrt{-3\Lambda}}{2}}
(1+\omega)t\Big)},
\end{equation}
where from (\ref{eq:solveldrho}) $\rho\propto R^{-3(1+\omega)}$. This is  an oscillating 
Universe with
singularities at $R=0$, $t=\frac{2\pi n}{\sqrt{-3\Lambda}(1+\omega)}$;
n is a half-integer. 

The partial differential equation  (\ref{eq:diffMpwrho}) for $M$ must also give these solutions. We now consider 
this equation for $\Lambda=0$.
Let us set $Y=2m+2M$. In terms of $Y$ equation (\ref{eq:diffMpwrho}) for an expanding Universe becomes

\begin{equation}\label{eq:diffY}
 R^{\frac{1}{2}}\frac{\partial Y}{\partial\tau}+(1+\omega)Y^{\frac{1}{2}}
 \frac{\partial Y}{\partial R}=0,
\end{equation}
which is easily solved by separation of variables. The solution is

\begin{equation}
A=1- \frac{2(m+M)}{R}=1-\frac{\bigg(\frac{2cR^{\frac{3}{2}}+3c_{2}}{3(1+\omega)(c t+c_{1})}\bigg)^{2}}{R},
\end{equation}
where $c$ is the separation constant and $c_{1}$ and $c_{2}$ are integration
constants. In the limit $R\rightarrow \infty$ the solution should give eq. (\ref{eq:Aflat}),
 this requires that $c_{1}=0$. Consider now the case where R
is finite and $t\rightarrow \infty$. We would in this case, for an embedded
mass, expect the solution to approach the Schwarzschild solution. But here the
solution goes towards Minkowski spacetime, indicating that our solution represents a pure
($m=0$) Universe. Thus, we should set $c_{2}=0$; then the
separation constant cancels and the solution reduces to   (\ref{eq:Aflat}).
This  method of solving
(\ref{eq:diffMpwrho}) cancels out the effect of m. But from this we see that for
$m=0$  equation
(\ref{eq:diffMpwrho}) gives the  required solution.

\subsection{$m\neq 0$}

We now turn to the case where we have an embedded mass, $m\neq 0$.
In \cite{gau2} no discussion was made on solutions of equation 
(\ref{eq:diffMpwrho}). We consider it for a flat Universe with 
$\Lambda=0$.

To find  solutions that include the effect of the embedded mass
m we shall look at approximations.  
We will find solutions valid close to 
m and at large distances from m.

We expand the square root
\begin{equation}\label{eq:Mmexp}
(2m +2M)^{\frac{1}{2}}=(2m)^{\frac{1}{2}}\Bigg(1+\frac{1}{2}\frac{M}{m}-
\frac{1}{2\cdot4}\bigg(\frac{M}{m}\bigg)^{2}+\ldots\Bigg),
\end{equation}
thus defining the function $F$:

\begin{equation}
(2m +2M)^{\frac{1}{2}}=(2m)^{\frac{1}{2}}+2F. 
\end{equation}
From (\ref{eq:diffMpwrho}) we get the following equation for $F$:

\begin{equation}\label{eq:diffF}
 R^{\frac{1}{2}}\frac{\partial F}{\partial\tau}+(1+\omega)(2m)^{\frac{1}{2}}
 \frac{\partial F}{\partial R}+
(1+\omega) 2F\frac{\partial F}{\partial R}=0,
\end{equation}
which has two approximations:
\begin{alignat}{2}\label{eq:appxF}
&R^{\frac{1}{2}}\frac{\partial F}{\partial\tau}+(1+\omega)(2m)^{\frac{1}{2}}
 \frac{\partial F}{\partial R}\approx 0,& \qquad 2F&\ll \left(2m\right)^{\frac{1}{2}},  \\
\nonumber\\
&R^{\frac{1}{2}}\frac{\partial F}{\partial\tau}+(1+\omega)
 2F\frac{\partial F}{\partial R}\approx 0,& \qquad 2F&\gg \left(2m\right)^{\frac{1}{2}}. \label{eq:appxFF}
\end{alignat}
Here (\ref{eq:appxF}) gives an approximate solution for the spacetime close to
the embedded mass m, and (\ref{eq:appxFF}) gives an approximate solution for
the spacetime far from $m$. Both equations are easily solved by separation of variables.

From equation  (\ref{eq:appxFF}) we find the following solution:

\begin{equation}\label{eq:findgeo}
(2m +2M)^{\frac{1}{2}}=(2m)^{\frac{1}{2}} +
\frac{2R^{\frac{3}{2}}}{3(1+\omega)t},
\end{equation}
where we have set the integration constants to zero.
The metric is given by

\begin{equation}
A=1-\Bigg(\bigg(\frac{2m}{R}\bigg)^{\frac{1}{2}} +\frac{2R}{3(1+\omega)t}\Bigg)^{2}.
\end{equation} 
This solution is valid for $2F\gg\left(2m\right)^{\frac{1}{2}} $ which is

\begin{equation}\label{eq:rangfar}
\frac{2R^{3}}{9(1+\omega)^{2}t^{2}}\gg m.
\end{equation}
i.e. where the Shcwarzschild field is much weaker than the background (see
(\ref{eq:Mmnull}) and (\ref{eq:rot})).
Taking the limit $R\rightarrow \infty$ this solution approaches the solution (\ref{eq:Aflat}) obtained for $m=0$.
Since the solution is only valid far from the embedded mass,
 M is now  not a global quantity; i.e. it does not represent the 
mass of the Universe.
From  (\ref{eq:partM}) we  find the density $\rho$
of the cosmic fluid with an embedded mass, 

\begin{equation}
\rho(t,R)=\frac{1}{(1+\omega)^{2}6\pi t^{2}}+
\frac{(2m)^{\frac{1}{2}}}{(1+\omega)4\pi R^{2}t}.
\end{equation}
The last term gives the deviation from the  density  (\ref{eq:rot}) 
of the cosmic fluid in a FRW Universe. 
We get the geodesics by inserting  (\ref{eq:findgeo}) into  
(\ref{eq:veldvacR}). Thus, far from m we have

\begin{equation}
HR=\frac{dR}{dt}=\left(\frac{2m}{R}\right)^{\frac{1}{2}}+
\frac{2R}{3(1+\omega)t}. 
\end{equation}
The solution for $\omega=0$ is 
\begin{equation}\label{eq:wzgeo}
R=\left(\frac{3}{2}\sqrt{2m}\ln{t}+c \right)^{\frac{2}{3}}t^{\frac{2}{3}}
\end{equation}
where $c$ is the integration constant. We note that for $t\rightarrow 0$
we get $R\rightarrow 0$. Using eq. (\ref{eq:rangfar}) we see that
this solution is valid for:

\begin{equation}
t\gg e^{p_{1}},\qquad p_{1}=1-\frac{2c}{3(2m)^{\frac{1}{2}}}
\end{equation}
and 
\begin{equation}
t\ll e^{p_{2}},\qquad p_{2}=-1-\frac{2c}{3(2m)^{\frac{1}{2}}}.
\end{equation}

For $\omega\neq0$ the solution is
\begin{equation}\label{eq:wgeo}
R=\left(\frac{3}{2}\sqrt{2m}\left(\frac{1}{\omega}+1\right)t+ c t^{\frac{1}{1+\omega}}\right)^{\frac{2}{3}}.
\end{equation}
Valid for

\begin{equation}
t\ll \left(\frac{2c}{3(2m)^{\frac{1}{2}}\left(\frac{\omega^{2}-\omega}
{1+\omega}\right)}\right)^{\frac{1+\omega}{\omega}}
\end{equation}
For $m=0$ the solutions (\ref{eq:wzgeo}) and 
(\ref{eq:wgeo})   reduces to eq.  (\ref{eq:soltra}) with the identification
$c=\left((1+\omega)\sqrt{6\pi b}\right)^{\frac{1}{\omega+1}}$.

Next consider the approximation (\ref{eq:appxF}). This equation has solution

\begin{equation}\label{eq:solmgg}
(2m +2M)^{\frac{1}{2}}=(2m)^{\frac{1}{2}} +
 ae^{b x},
\end{equation}
where 
\begin{equation}
x=\frac{2R^{\frac{3}{2}}}{3(1+\omega)(2m)^{\frac{1}{2}}}-t,
\end{equation} 
$a$ is an integration constant and $b$ is the separation constant.
This solution is valid for
\begin{equation}\label{eq:rangclos}
bx\ll \ln{\frac{(2m)^{\frac{1}{2}}}{a}}.
\end{equation}

The metric function A  becomes
\begin{equation}
A=1-\frac{\left((2m)^{\frac{1}{2}} + ae^{ b x}\right)^{2}}{R}.
\end{equation}
For a finite R in the limit $t\rightarrow \infty$ we get
$A\rightarrow 1-\frac{2m}{R}$, i.e. the solution approaches the Schwarzschild solution. Also, for $a=0$, which gives $M=0$, the second term is zero and we are left with the Schwarzschild field.
Considering (\ref{eq:solmgg}) to first order in $\frac{M}{m}$
this solution reduces to

\begin{equation}\label{eq:appMmfs}
M=(2m)^{\frac{1}{2}}ae^{bx}.
\end{equation} 
If we assume  $M \ll m$ in (\ref{eq:diffMpwrho})  
 and  ignore the non-linear term completely we can write
$R^{\frac{1}{2}}\frac{\partial M}{\partial t}+(2m)^{\frac{1}{2}}
 \frac{\partial M}{\partial R}=0$ which has solution (\ref{eq:appMmfs}).
From (\ref{eq:solmgg}) and (\ref{eq:veldvacR}) we have the geodesics close to 
$m$ approximated by

\begin{equation}\label{eq:geoexp}
HR=\frac{dR}{dt}=\left(\frac{2m}{R}\right)^{\frac{1}{2}}+ 
\frac{ae^{bx}}{R^{\frac{1}{2}}}. 
\end{equation}
If we write this in terms of $x$ and then insert
$Z=\frac{2F}{a}=e^{bx}$, eq. (\ref{eq:geoexp}) can be written
\begin{equation}
\frac{dZ}{dt}=\frac{ab Z^{2}-b\omega(2m)^{\frac{1}{2}}Z}{(1+\omega)
(2m)^{\frac{1}{2}}}, 
\end{equation}
which is easily integrated to find the trajectories.
For dust, $\omega=0$, the geodesics are 

\begin{equation}\label{eq:wmzgeo}
R(t)=\left(\frac{3(2m)^{\frac{1}{2}}}{2}t-
\frac{3(2m)^{\frac{1}{2}}}{2b}ln\left({c_{1}-\frac{ab}
{(2m)^{\frac{1}{2}}}t}\right)\right)^{\frac{2}{3}},
\end{equation}
where $c_{1}$ is the integration constant. If we demand that $R(t\!\!=\!\!0)\!=\!0$ we get
 $c_{1}=1$. From eq. (\ref{eq:rangclos}) we find that this solution is valid for

\begin{equation}
t\ll \frac{1}{b}\left(\frac{(2m)^{\frac{1}{2}}c_{1}}{a}-1\right).
\end{equation}
Thus, for this to be valid in an expanding Universe we must have
$\sqrt{2m}\gg a$.

For $\omega\neq 0$ the solution is

\begin{equation}\label{eq:wmgeo}
R(t)=\left(\frac{3(2m)^{\frac{1}{2}}}{2}t + \frac{3(1+\omega)
(2m)^{\frac{1}{2}}}{2b}
ln\left({\frac{\omega(2m)^{\frac{1}{2}}}{ac_{2}+ae^{-\frac{\omega}
{1+\omega}bt}}}\right)\right)^{\frac{2}{3}}.
\end{equation}
Putting $R(t\!\!=\!\!0)\!=\!0$ gives the integration constant $c_{2}=
\frac{\omega (2m)^{\frac{1}{2}}}{a}-1$. This solution is valid for

\begin{equation}
e^{\frac{\omega}{1+\omega}bt}\gg \frac{\omega-1}{c_{2}}.
\end{equation}
By setting $a=0$ in (\ref{eq:wmzgeo}), or (\ref{eq:wmgeo}),
we are left with the radial Schwarzschild geodesics.

\section{Singular shells in isotropic Universe models}\label{ch:isformgen}
We are considering a Universe containing, in addition to the cosmic fluid and $\Lambda$, 
energy confined to a surface (or rather an hypersurface). That is, we are dealing
with situations where we have a shell of energy where the thickness, $\vartheta$,  of the shell can be ignored; i.e. mathematically we let the thickness go to zero,
$\vartheta\rightarrow 0$. This is the thin shell approximation.
The energy momentum tensor for such a spacetime can be split into three parts.
A general $T_{\alpha\beta}$  can therefore be written

\begin{equation}\label{eq:TempmS}
T_{\alpha\beta}=S_{\alpha\beta}\delta(y)+T^{+}_{\alpha\beta}\theta(y)
+T^{-}_{\alpha\beta}\theta(-y),
\end{equation}\\
here $y$ is an orthogonal coordinate such that $\frac{\partial}{\partial y}=\mathbf n$
is a normalized normal vector  and  $y=0$ at the hypersurface.
$\delta(y)$ is a delta function and $\theta(y)$
is the step function. In the previous section we discussed the embedding of a 
 Schwarzschild mass into a cosmological model. Now we wish to obtain the relativistic equation of motion
for a thin  shell in this ambient space-time. From   (\ref{eq:TempmS}) we see that the shell contributes with a delta function singularity to the energy momentum tensor and thus does not follow geodesics in the background spacetime. In  the thin shell approximation
the energy-momentum tensor $S_{\alpha\beta}$ of the surface is defined as
the integral over  the thickness of the surface when the thickness $\vartheta$ goes to zero 

\begin{equation}
S_{\alpha\beta}=\lim_{\vartheta\rightarrow 0}\int_{-\frac{\vartheta}{2}}
^{\frac{\vartheta}{2}}T_{\alpha\beta}dy.
\end{equation}

\subsection{Israel's Formalism: The metric junction method}
To deal with the situation described above we use Israel's formalism 
\cite{is1}. In this section we give a review.

The spacetime manifold $\mathcal{M}$  is split into two parts, 
$\mathcal{M}^{+}$ and $\mathcal{M}^{-}$, by a hypersurface $\Sigma$.    
That is, $\mathcal{M}^{+} \cup \mathcal{M}^{-}=\mathcal{M}$ with a common boundary: $\partial\mathcal{M}^{+} \cap \partial\mathcal{M}^{-}=\Sigma$. 

The energy-momentum content of spacetime is coupled to the geometry through
Einstein's field equations. In the regions 
$\mathcal{M}^{+}$
and $\mathcal{M}^{-}$ outside the hypersurface $\Sigma$ we assume that

\begin{equation} \label{eq:ein+-}
G^{\pm}_{\mu\nu}= 8\pi T^{\pm}_{\mu\nu},
\end{equation}
where $+$ and $-$ means the tensor evaluated in $\mathcal{M}^{+}$
and $\mathcal{M}^{-}$, respectively.
Thus, if $T^{\pm}_{\mu\nu}$ is given, the metrics for the two regions 
outside  $\Sigma$ are obtained by solving (\ref{eq:ein+-}).
The line elements of the two regions are written
\begin{equation}
ds^{2}_{\pm}=g_{\mu\nu}^{\pm}dx^{\mu}_{\pm}dx^{\nu}_{\pm}.
\end{equation}
On  the hypersurface $\Sigma$ we denote the intrinsic coordinates by $\xi^{j}$,
and the intrinsic (induced) metric is

\begin{equation}
ds^{2}_{\Sigma}=h_{ij}d\xi^{i}d\xi^{j}.
\end{equation}
We use Greek indices to run over the coordinates  of 
$\mathcal{M}^{\pm}$ ($d$ dimensional) while Latin indices run over the 
intrinsic coordinates  of $\Sigma$ ($d-1$ dimensional).

The equations for the hypersurface are given by the embeddings $\phi^{\pm}$:

\begin{equation}
x^{\mu}_{\pm}=\phi^{\mu}_{\pm}(\xi^{i})
\end{equation}
where $x^{\mu}_{\pm}$ are $c^{r}$ functions with 
$r\geq 4$.\footnote{The continuity equations involves the 4th
derivative of  $x^{\mu}_{\pm}$.}
To glue together $\mathcal{M}^{+}$ and $\mathcal{M}^{-}$ along the common boundary
$\Sigma$ we require that the two metrics, $g^{+}_{\mu\nu}$ and $g^{-}_{\mu\nu}$,
 induce the same intrinsic metric on $\Sigma$:

\begin{equation}\label{eq:metjun}
h_{ij}= g_{\mu\nu}^{+}\frac{\partial x^{\mu}_{+}}{\partial \xi^{i}}
                       \frac{\partial x^{\nu}_{+}}{\partial \xi^{j}}
      = g_{\mu\nu}^{-}\frac{\partial x^{\mu}_{-}}{\partial \xi^{i}}
                       \frac{\partial x^{\nu}_{-}}{\partial \xi^{j}}.	
\end{equation}\\
We note that the junction is independent of the embeddings $x^{\mu}_{\pm}$, which do not need to join continuously at the hypersurface. This is an essential property of Israel's formalism: We are free to choose coordinates in $\mathcal{M}^{+}$ and $\mathcal{M}^{-}$,
independently.

Let $\mathbf{n}$ be the unit normal to the hypersurface $\Sigma$, which
is defined to  point from
$\mathcal{M}^{-}$ to $\mathcal{M}^{+}$, such that:

\begin{eqnarray}
 \mathbf{n} \cdot \mathbf{n} =g_{\mu\nu}n^{\mu}n^{\nu}\mid^{\pm}=\epsilon= \left\{
 \begin{array}{c}
 +1 \\ -1
 \end{array} \right.
\end{eqnarray}\\
$\epsilon=+1$ gives a spacelike $\mathbf{n}$ and thus a timelike hypersurface
$\Sigma$. $\epsilon=-1$ gives a timelike $\mathbf{n}$ and thus a spacelike 
$\Sigma$. For $\epsilon=0$ we call the hypersurface a null surface\footnote{The
formalism given here breaks down for this case, for a treatment of null surfaces see \cite{bais,be2,cldr}.}. 
We shall only consider timelike surfaces, $\epsilon=+1$.

The induced metric on $\Sigma$ can also be written in terms of $x_{\pm}^{\mu}$. 
We have $h_{\pm}^{\mu\nu}= 
h^{ij}\frac{\partial x^{\mu}_{\pm}}{\partial \xi^{i}}
\frac{\partial x^{\nu}_{\pm}}{\partial\xi^{j}}$. 
In terms of the normal vector the induced metric becomes

\begin{equation}\label{eq:hhyp}
h^{\mu\nu}_{\pm}=
g^{\mu\nu}_{\pm} -\epsilon n^{\mu}_{\pm}n^{\nu}_{\pm},
\end{equation}
$h_{\pm}^{\mu\nu}$ defines a projection operator, 
$h_{\alpha}^{\mu}h^{\alpha}_{\nu}=h_{\nu}^{\mu} $, 
that picks out the part of a tensor that lies in the tangent space of $\Sigma$.   

The extrinsic curvature tensor, $\mathbf K$, is essential to the formalism.  
This tensor is a measure of how the hypersurface curves in the surrounding
spacetime $\mathcal{M}^{+}$ and $\mathcal{M}^{-}$. 
It is defined as the covariant derivative of the normal vector $\mathbf{n}$ with respect to the
connection in $\mathcal{M}^{+}$ and $\mathcal{M}^{-}$  along the direction of the tangent vectors on $\Sigma$.
Since $\mathbf{n}$ is everywhere normal to $\Sigma$ and of constant magnitude, the variation of $\mathbf{n}$ and thus   $\mathbf K$ 
will be entirely in the tangent space of $\Sigma$. Also, since there is a delta function singularity in the energy momentum tensor
 at $\Sigma$, 
we get a discontinuity in   $\mathbf K$ at $\Sigma$ ($\delta(y)=\theta'(y)$).   
The extrinsic curvature tensors 
$K_{ij}^{+}$ and $K_{ij}^{-}$ in 
$\mathcal{M}^{+}$ and $\mathcal{M}^{-}$, respectively, are thus given by

\begin{equation}\label{eq:extensor}
K_{ij}^{\pm}=- 
\frac{\partial x^{\alpha}}{\partial \xi^{i}}\mathbf{e}_{\alpha}\cdot 
\frac{\partial x^{\beta}}{\partial \xi^{j}}\nabla_{\beta} \mathbf{n}\Big|^{\pm}.
\end{equation}
Note the sign convention.  We can express $K_{ij}$ in terms of the Christoffel symbols
of $\mathcal{M}^{\pm}$. Using $\nabla({\mathbf{n}\cdot
\frac{\partial x^{\alpha}}{\partial \xi^{i}}\mathbf{e}_{\alpha}})=0$ we have

\begin{equation}\label{eq:exchcon}
K_{ij}^{\pm}=
\mathbf{n}\cdot \frac{\partial x^{\alpha}}{\partial \xi^{i}}
\nabla_{\alpha}\bigg(\frac{\partial x^{\beta}}{\partial \xi^{j}}
\mathbf{e}_{\beta}\bigg)\bigg|^{\pm}=
\bigg(\frac{\partial^{2}x^{\mu}}{\partial\xi^{i}\partial\xi^{j}}+
\Gamma^{\mu}_{~\alpha\beta}
\frac{\partial x^{\alpha}}{\partial \xi^{i}}
\frac{\partial x^{\beta}}{\partial \xi^{j}}
\bigg)n_{\mu} \Big|^{\pm}.
\vspace*{3mm}%
\end{equation}
We see that $K_{ij}$ is symmetric and represents 4-scalars in 
$\mathcal{M}^{\pm}$. In terms of the projection operator (\ref{eq:hhyp}) we have
$K_{\mu\nu}^{\pm}=-h^{\alpha}_{\mu}h^{\beta}_{\nu}n_{\alpha;\beta}|^{\pm}$ and 
$K_{ij}^{\pm}= K_{\mu\nu}\frac{\partial x^{\mu}}{\partial \xi^{i}}\frac{\partial x^{\nu}}{\partial \xi^{j}}|^{\pm}$.

In Gaussian normal coordinates  the metric is given by
\begin{equation} 
ds^{2}= dn^{2}+ h_{ij}(\xi^{i},n)d\xi^{i}d\xi{j}, 
\end{equation}
where $\Sigma$ is located at $n=0$ and the induced metric on  $\Sigma$ is
$h_{ij}(\xi^{i},0)= h_{ij}(\xi^{i})$. In this case
the extrinsic curvature tensor is simply given by
$K_{ij}=\Gamma^{n}_{~ij}=-\frac{1}{2}h_{ij,n}$ and
$K^{i}_{~j}=-\Gamma^{i}_{~nj}=-h^{ik}\frac{1}{2}h_{kj,n}$.

The Einstein equations  on the hypersurface are:

\begin{eqnarray}
-\frac{1}{2}\epsilon \; ^{(3)}R + 
\frac{1}{2}(K^{2}- K_{lm}K^{lm})\!\mid^{\pm} &=&8\pi T_{nn}^{\pm} 
                                           \label{eq:eqeinnor}      \\ 
                                                           \nonumber \\
 -(K_{i~~|l}^{~l}-K_{|i})\!\mid^{\pm}&=&8\pi T_{in}^{\pm}
                                             \label{eq:eqeinmix}      \\
                                           \nonumber \\
\; ^{(3)}G_{ij}+\epsilon(K_{ij}-g_{ij}K)_{,n}\!\mid^{\pm}
-3\epsilon K_{ij}K\!\mid^{\pm}&&                         \nonumber  \\
+2\epsilon K_{i}^{~l}K_{jl}\!\mid^{\pm} 
 +\epsilon \frac{1}{2}g_{ij}(K^{2} + K_{lm}K^{lm})\!\mid^{\pm}&=&
8\pi T_{ij}^{\pm}, \label{eq:eqeintan}
 \end{eqnarray}

Integrating (\ref{eq:eqeinnor}) and (\ref{eq:eqeinmix}) across the shell in 
the thin shell approximation we find $S_{in}=S_{nn}=0$. That is, 
$\mathbf S$ has no normal components to the shell. We have
$S_{\mu\nu}^{\pm}=h^{\alpha}_{\mu}h^{\beta}_{\nu}S_{\alpha\beta}|^{\pm}$ and 

\begin{equation}\label{eq:s3s4}
S_{ij}=\frac{\partial x^{\alpha}_{\pm}}{\partial \xi^{i}}
       \frac{\partial x^{\beta}_{\pm}}{\partial\xi^{j}}
       S_{\alpha\beta}\big|^{\pm}.
\end{equation}

Integrating  (\ref{eq:eqeintan}) we arrive at the equation of motion  for the surface:

\begin{equation}\label{eq:lanczos}
[K_{ij}]-[K]g_{ij}=8\pi S_{ij}.
\end{equation}\\
By contraction 
\begin{equation}\label{eq:conlanczos}
[K_{ij}]=8\pi (S_{ij}-\frac{1}{2}Sg_{ij}).
\end{equation}\\ 
These equations  are called the Lanczos equations and they
 say that the surface energy-momentum tensor is given by the difference 
in the embeddings of $\Sigma$ in $\mathcal{M}^{\pm}$.
The bracket operation $[\;]$ gives the discontinuity of a tensor at $\Sigma$ 
\begin{equation}
[T]=T^{+}-T^{-}.
\end{equation}
In the same way we define
the average $\{\;\}$ as
\begin{equation}
\{T\}=\frac{1}{2}(T^{+}+T^{-}).
\end{equation}
We also note two relations between these  definitions:

\begin{equation}\label{eq:sym}
[TS]=[T]\{S\}+\{T\}[S]
\end{equation}
\begin{equation}\label{eq:asym}
\{TS\}=\{T\}\{S\}+\frac{1}{4}[T][S].
\end{equation}
\vspace{0.5mm}

Using $[\;]$ on the Einstein equations (\ref{eq:eqeinnor}) and 
(\ref{eq:eqeinmix}) along with the Lanczos equation (\ref{eq:lanczos}) and
equation (\ref{eq:sym}) we find: 

\begin{equation}\label{eq:junnn}
S_{lm}\{K^{lm}\}+[T_{nn}]=0.
\end{equation}\\
The discontinuity $[T_{nn}]$ gives the pressure exerted normal to the surface by the bulk.
Contracting eq. (\ref{eq:conlanczos}) with $S_{ij}$ we get
\begin{equation}\label{eq:junnncon}
S_{lm}[K^{lm}]=8\pi (S_{ij}S^{ij}-\frac{1}{2}S^{2}).
\end{equation}
The contraction  $S_{lm}K^{lm}_{\pm}$ gives the normal component of the divergence of $\mathbf{S}$ with respect to the connection on $\mathcal{M}^{\pm}$.

\begin{equation}
S^{\alpha\beta}_{~~~;\beta}\!\mid^{\pm}=
\left(\frac{\partial \phi^{\alpha}}{\partial \xi^{i}}S_{~~|l}^{il}+
S_{lm}K^{lm}n^{\alpha}\right)\!\Big|^{\pm},
\end{equation}
where $|$ denotes 
covariant derivative with respect to the intrinsic connection on $\Sigma$
(i.e. the metric connection defined by $h_{ij}$).
The normal part is

\begin{equation}\label{eq:nordivs}
n_{\alpha}S^{\alpha\beta}_{~~~;\beta}\!\mid^{\pm}=
S_{lm}K^{lm}_{\pm}.
\end{equation}

Combining   (\ref{eq:junnn}) and (\ref{eq:junnncon}) we can separate 
$S_{lm}K^{lm}_{+}$ and  $S_{lm}K^{lm}_{-}$
\cite{be2}. 
In  a spherically symmetric space-time with the surface consisting of an ideal fluid we may separate  the time components of $K^{lm}_{+}$ and  $K^{lm}_{-}$.

The continuity equation for the surface is

\begin{equation}\label{eq:junin}
S_{i~~|l}^{~l}+[T_{in}]=0.
\end{equation}
Contracting $[T_{in}]$ with $u^{i}$  we get the momentum-flux of the bulk as measured by a comoving observer to the surface. Contracting with a spacelike tangent vector, $x^{i}$, this term  gives the tangential force, in the
$x^{i}$ direction, exerted on the surface by the bulk.

Taking the covariant derivative of the 4-velocity $\mathbf{u}_{\pm}$ in the direction of
$\mathbf{u}_{\pm}$ with respect to the connection on $\mathcal{M}^{\pm}$ we obtain 
the  4-acceleration $\mathbf{a}_{\pm}$. As viewed from $\mathcal{M}^{\pm}$

\begin{eqnarray}
a^{\alpha}\vec{e}_{\alpha}\!\mid^{\pm}=\nabla_{\vec{u}}\vec{u}\!\mid^{\pm}
=u^{l}_{~|m}u^{m}\vec{e}_{l}+K_{lm}u^{l}u^{m}\vec{n}|^{\pm}.
\end{eqnarray}
The normal component of the  4-acceleration 
of the shell as viewed from $\mathcal{M}^{\pm}$ is thus

\begin{equation}\label{eq:accexcurv}
a^{\alpha}n_{\alpha}\!\mid^{\pm}=K_{lm}^{\pm}u^{l}u^{m},
\end{equation}
which is in general non-zero. 

Applying $\{\;\}$ on (\ref{eq:eqeinnor}) and 
(\ref{eq:eqeinmix}) gives the two constraints:
\begin{eqnarray}
&&^{(3)}R -\{K^{2}\}+\{K_{lm}\}\{K^{lm}\}=-16\pi^{2}(S_{lm}S^{lm}+S^{2})-
16\pi\{T_{nn}\}\;\;\\
&&\{K^{~l}_{i~|l}\}-\{K_{,i}\} =-8\pi\{T_{in}\}.
\end{eqnarray}

\subsection{Spherically symmetric thin shells}
We will look at a spherically symmetric shell consisting of an ideal
fluid.  The spherical symmetry implies that we can write the line-element,
using proper time on the shell, as:
 
\begin{equation}\label{eq:intmet}
ds^{2}_{_{\Sigma}}= -d\tau^{2}+R(\tau)^{2}d\Omega^{2}
\end{equation}
where $R(\tau)$ is the expansion factor for the hypersurface,
and  the proper area is given by
\begin{equation}
A_{\Sigma}=4\pi R^{2}.
\end{equation}

The energy-momentum tensor for the shell consisting of an ideal fluid is

\begin{equation}\label{eq:intSif}
S_{ij}=(\sigma+p)u_{i}u_{j} +p\;g_{ij},
\end{equation}
where $\sigma$ is the mass (energy) density of the surface
 and $p$ is the tangential pressure of
the surface. Since we use proper
time the 4-velocity of a comoving observer is $u^{i}=(1,0,0)$.
The  mass, $\mu$, of the surface is

\begin{equation}
\mu=4\pi R^{2} \sigma.
\end{equation}

The angular coordinates define tangent vectors to the surface. Thus the
radial coordinates must join continuously on the hypersurface. Hence, 
the equation of the surface is given by
\begin{equation}
 R_{\pm}=R(\tau),
\end{equation}
henceforth omitting the subscripts  on R.
The line element in the  spacetimes
 $\mathcal{M}^{+}$ and $\mathcal{M}^{-}$ when 
including the effect of a Schwarzschild mass on the background is given in
 (\ref{eq:metradgeot}). That is,

\begin{eqnarray}\label{eq:metEdSSjun}
ds_{\pm}^{2}&=&\mathcal E_{\pm}^{-2}\left(
-A_{\pm}dt_{\pm}^{2}-2(1-A_{\pm})^{\frac{1}{2}}
   dt_{\pm}dR +dR^{2}\right)+ R^{2}d\Omega^{2} \nonumber \\
  &=& -dt_{\pm}^{2} +\mathcal E_{\pm}^{-2}\Big(dR-(1-A_{\pm})^{\frac{1}{2}}
   dt_{\pm}\Big)^{2}+ R^{2}d\Omega^{2},
\end{eqnarray}
where in the most general case:

\begin{eqnarray}\label{eq:Apg}
&&A_{+}=1-\frac{2(m_{+}+M_{+})}{R}-\frac{\Lambda_{+}}{3}R^{2}\\
 \nonumber \\
&&A_{-}=1-\frac{2(m_{-}+M_{-})}{R}-\frac{\Lambda_{-}}{3}R^{2},\label{eq:Amg}
 \end{eqnarray}
$m_{-}$ is the Schwarzschild mass at the center, and
$m_{+}=m_{-}+m_{_{\Sigma}}$ where
$m_{_{\Sigma}}$ is the Schwarzschild mass of the shell;  $\Lambda_{\pm}$  gives
the vacuum energy density in $\mathcal{M}^{\pm}$, respectively; 
$M_{\pm}$ are the mass functions for
the Universe in $\mathcal{M}^{\pm}$, respectively,  defined in
equation (\ref{eq:M}). The  parameter $\mathcal E_{\pm}$ is given in eq.
(\ref{eq:Etop}).

The  embedding is

\begin{equation}
x^{\mu}_{\pm}=(t_{\pm}(\tau), R(\tau), 0, 0).
\end{equation}
Thus, the 4-velocity, $\mathbf{u}_{\pm}$,    for a comoving observer is
  
\begin{equation}
\frac{\partial x^{\mu}_{\pm}}{\partial\tau}=\frac{d x^{\mu}_{\pm}}{d\tau}
\equiv u^{\mu}=(\dot{t}_{\pm},\dot{R}(\tau), 0, 0),
\end{equation}
where the dot denotes derivative with respect to the proper time $\tau$ 
on the shell. In terms of the intrinsic coordinates we have

\begin{equation}\label{eq:pushu}
u^{i}\frac{\partial \phi^{\alpha}_{\pm}}{\partial \xi^{i}}=u^{\alpha}_{\pm},
\end{equation}
i.e. $u^{i}=(1,0,0)$.

From $n^{\alpha}n_{\alpha}|^{\pm}=1$   and           
$u^{\alpha}n_{\alpha}|^{\pm}=0$ using $u^{\alpha}u_{\alpha}|^{\pm}=-1$ 
we  find the following covariant components for the normal vector:

\begin{equation}\label{eq:nor}
n_{\mu}^{\pm}=-\zeta_{\pm}\mathcal E_{\pm}^{-1}(-\dot{R},\dot{t}_{\pm},0,0),
\end{equation}
where $\zeta=\pm 1$.

From the $h_{\tau\tau}$ component of the metric junction 
(\ref{eq:metjun}), (or from the line-element: $ds_{\pm}^{2}=-d\tau^{2}$,  
$\tau$ is proper time on the shell) we have:

\begin{equation}\label{eq:ttjun}
-A_{\pm}\dot{t}^{2}_{\pm}-2(1-A_{\pm})^{\frac{1}{2}}\dot{t}\dot{R}
+\mathcal E_{\pm}^{2} +\dot{R}^{2}=0
\end{equation}
which is a quadratic equation in $\dot{t}_{\pm}$. Solving this gives:

\begin{equation}\label{eq:dottau}
\dot{t}_{\pm}=\frac{(\mathcal E_{\pm}^{2}-A_{\pm})^{\frac{1}{2}}
\dot{R} - \mathcal E_{\pm}\sqrt{A_{\pm}+\dot{R}^{2}}}{-A_{\pm}}.
\end{equation}
We  make the sign choice by requiring that  
 $\tau$ and $t_{\pm}$ are pointing in the same direction: 
$\dot{t}_{\pm}>0$. Eq. (\ref{eq:dottau}) can be rearranged to give

\begin{equation}\label{eq:dottauto}
\dot{t}_{\pm}=\frac{\mathcal E_{\pm}^{2}+\dot{R}^{2}}
{(1-A_{\pm})^{\frac{1}{2}}\dot{R} + \mathcal E_{\pm}\sqrt{A_{\pm}+
\dot{R}^{2}}},
\end{equation}
We see that this is well behaved at the
horizons $A=0$ and crossing the horizon such that $A<0$ we cannot have
stationary shells, $\dot{R}\neq0$. 
Also, we note here that $t_{\pm}$ 
cannot in general  join
continuously at the surface when embedding $m_{_{\Sigma}}$ in 
$\mathcal{M}^{+}$. Eq. (\ref{eq:ttjun}) also gives

\begin{equation}\label{eq:covelgsh}
\frac{dR}{dt_{\pm}}=(\mathcal E_{\pm}^{2}-A_{\pm})^{\frac{1}{2}}\pm 
\mathcal E_{\pm} (1-\dot{t}_{\pm}^{-2})^{\frac{1}{2}}=H_{\pm}R+v_{\pm}.
\end{equation}
From (\ref{eq:velgenR}) we see that the first term gives the expansion of the 
Universe. Thus, the second term
represents the velocity, $v_{\pm}$, of the shell relative to the expansion in
$\mathcal M^{+}$ and $\mathcal M^{-}$, respectively. Then $\dot{t}_{\pm}$ is given by the Lorentz factor

\begin{equation}\label{eq:dtlo}
\dot{t}_{\pm}=\frac{1}{\sqrt{1-\frac{v_{\pm}}{\mathcal E_{\pm}}}}.
\end{equation}

From eq. (\ref{eq:exchcon})  we find the angular component of the extrinsic curvature tensor, 
\begin{equation}\label{eq:Kang}
K_{\theta\theta}=-Rn^{R}=\zeta_{\pm}R\sqrt{A_{\pm}+\dot{R}^{2}}, 
\end{equation}
where we have used (\ref{eq:nor}) and (\ref{eq:dottau}).
Thus, from Lanczos eq. (\ref{eq:conlanczos}) we get the equation of 
motion:

\begin{equation}\label{eq:geneqm}
\zeta_{+}\sqrt{A_{+}+\dot{R}^{2}} -\zeta_{-} \sqrt{A_{-}+\dot{R}^{2}}=\frac{\mu}{R}=4\pi R\sigma.
\end{equation}
This equation  generalizes the previous  junctions to include the
junction for a general Universe with an embedded Schwarzschild mass.

In terms of the Gaussian normal coordinates we have 
$K_{\theta\theta}=-R\frac{\partial R}{\partial n}$; i.e. 
 the sign $K_{\theta\theta}$ determines whether the radius of the surface is 
increasing or decreasing in the normal direction. For static spacetimes 
 the sign of $K_{\theta\theta}$ is given by
$\zeta$ and  determines the spatial topology\cite{be1,guth,be2,sato}.
If either of the spacetimes are not static  the sign of $K_{\theta\theta}$
does not in general give the topology;
for the classification of the junction of two FRW spacetimes see \cite{sama1}. 

Squaring (\ref{eq:geneqm}) we find
\begin{equation}\label{eq:sqeqm}
\zeta_{\pm}\sqrt{A_{\pm}+\dot{R}^{2}}= 
\frac{R}{2\mu}(A_{+}-A_{-})\pm\frac{\mu}{2R}
\end{equation}
from which one can determine the sign of $K_{\theta\theta}$.
Squaring (\ref{eq:sqeqm})  we get the energy equation 

\begin{equation}\label{eq:Rdot}
\dot{R}^{2}=\dot{t}_{\pm}^{2}\left(\frac{dR}{dt_{\pm}}\right)^{2}=\frac{\mu^{2}}{4R^{2}}-\frac{1}{2}(A_{+}+A_{-})
+\frac{R^{2}}{4\mu^{2}} (A_{+}-A_{-})^{2}.
\end{equation}

The continuity equation (\ref{eq:junin})  gives for a comoving observer:

\begin{equation}
\dot{\sigma}=-2(\sigma+p)\frac{\dot{R}}{R}+[T_{\mu\nu}n^{\mu}u^{\nu}],
\end{equation}
or in terms of $\mu$ we have
\begin{equation}\label{eq:dotmu}
\dot{\mu}=-8\pi p R \dot{R}+4\pi R^{2}[T_{\mu\nu}n^{\mu}u^{\nu}].
\end{equation}
The first term is due to the
tangential pressure of the surface. The second term, 
the area of the shell times the discontinuity in
the 4-momentum,
 can be interpreted as the 
mass gathered on the shell from the surroundings due to the motion of 
the shell 
relative to the cosmic fluid. 
The force needed to move on the surface is given by
\begin{equation}
(\sigma+p)u_{i\mid l}u^{l}=-(\delta^{l}_{~i}+u_{i}u^{l})p_{,l}
-u_{i}[T_{\mu\nu}n^{\mu}u^{\nu}]-
[T_{\mu\nu}n^{\mu}\frac{\partial \phi^{\nu}}{\partial \xi^{i}}],
\end{equation}
which for a comoving observer is zero, i.e. following the intrinsic geodesics of the surface.

The energy-momentum  tensor is given in 
equation (\ref{eq:cosflu}), where $V^{\mu}$ (eq. (\ref{eq:radgeot}))
is the  4-velocity of the cosmic fluid. Contracting with the normal 
$n_{\mu}$ and the
 4-velocity $u^{\mu}$ of the surface gives:
 $V^{\mu}n_{\mu}|^{\pm}=-\theta_{\pm}\left(
(1-A_{\pm})^{\frac{1}{2}}\dot{t}_{\pm}-\dot{R}\right)$, 
 and  $V^{\mu}u_{\mu}|^{\pm}=-\dot{t}_{\pm}$. Thus, when including the 
 vacuum energy, the components of the energy-momentum tensor at the surface
becomes: 
  
\begin{eqnarray}\label{eq:tnnemgen}
T_{\mu\nu}n^{\mu}n^{\nu}\mid^{\pm}&=&(\rho_{\pm}+p_{cf}^{\pm})\Big(\dot{R}
-(1-A_{\pm})^{\frac{1}{2}}\dot{t}_{\pm}\Big)^{2}+p_{cf}^{\pm}-\frac{\Lambda_{\pm}}
{8\pi} \\ 
\nonumber \\
T_{\mu\nu}n^{\mu}u^{\nu}\mid^{\pm}&=&
-\zeta_{\pm} (\rho_{\pm}+p_{cf}^{\pm})\dot{t}_{\pm}
\Big( \dot{R}-(1-A_{\pm})^{\frac{1}{2}}\dot{t}_{\pm}\Big), \label{eq:tnuemgen} 
\end{eqnarray}
where $p_{cf}$ is the pressure of the cosmic fluid and $\rho$ the energy density.
For a shell moving with the expansion of the Universe we have
$\dot{R}=(\mathcal E^{2}-A_{\pm})^{\frac{1}{2}}$ and $\dot{t}_{\pm}=1$.
This gives $T_{\mu\nu}n^{\mu}n^{\nu}\mid^{\pm}=p_{cf}^{\pm}
-\frac{\Lambda_{\pm}}{8\pi}$ 
and $T_{\mu\nu}n^{\mu}u^{\nu}\mid^{\pm}=0$.

For an equation of state $p_{cf}=\omega\rho$, $M$ satisfies eq.
(\ref{eq:diffMpwrho}). Thus, the derivative of $M$ with respect to the proper time $\tau$
on the surface may be written:

\begin{eqnarray} 
\dot{M}_{\pm}&=&\frac{\partial M_{\pm}}{\partial t_{\pm}}\left(
\dot{t}_{\pm}-(1+\omega_{\pm})(\mathcal E_{\pm}^{2}-A_{\pm})^{-\frac{1}{2}}\dot{R}\right)\\
\nonumber \\ 
\dot{M}_{\pm}&=&\frac{\partial M_{\pm}}{\partial  R}\left(
\dot{R}-(1+\omega_{\pm})(\mathcal E_{\pm}^{2}-A_{\pm})^{\frac{1}{2}}\dot{t}_{\pm}\right).\label{eq:dotMdmdr} 
\end{eqnarray}
We will in the following restrict our discussion to $\omega=0$.
We see that if the shell is moving with the expansion this is zero,
$\dot{M}=0$. Also, 
we find that $\dot{M}>0$ for a shell expanding faster than the Universe
and $\dot{M}<0$ if the shell is moving slower than the expansion of the
Universe. Using eq. (\ref{eq:dotMdmdr}) with $\omega=0$ the normal and mixed components of the energy momentum tensor (\ref{eq:tnnemgen}) and
(\ref{eq:tnuemgen}) can be written:
\begin{eqnarray}\label{eq:tnndotM}
T_{\mu\nu}n^{\mu}n^{\nu}\mid^{\pm}&=&\frac{\dot{M}_{\pm}^{2}}{4\pi R^{2}
 \frac{\partial M_{\pm}}{\partial  R}}-\frac{\Lambda_{\pm}}{8\pi}\\
\nonumber \\
T_{\mu\nu}n^{\mu}u^{\nu}\mid^{\pm}&=&-\zeta_{\pm}
 \frac{\dot{M}_{\pm}\dot{t}_{\pm}}
{4\pi R^{2}}.
\end{eqnarray}
Furthermore, $\dot{M}$ can be written in terms of $v_{\pm}$ and $\rho_{\pm}$
by using equations (\ref{eq:partM}), (\ref{eq:diffMpwrho}), 
(\ref{eq:covelgsh}) and (\ref{eq:dtlo}):
\begin{equation}
\dot{M}_{\pm}=\frac{v_{\pm}4\pi R^{2}\rho_{\pm}}{\sqrt{1-\frac{v_{\pm}}{\mathcal E_{\pm}}}}.
\end{equation}
This gives the rate of change of the  mass $\mu$,

\begin{equation}\label{eq:dotmuMpm}
\dot{\mu}=-8\pi p R\dot{R} -[\dot{t}v4\pi R^{2}\rho]
\end{equation}
where we have set $\zeta_{\pm}=-1$. For a surface with
 vanishing tangential pressure, $p=0$, this has the same form as eq.
2.13 in \cite{samasa}. 

Consider now the time component of the field equations for the surface.
Combining equations  (\ref{eq:junnn}) and (\ref{eq:junnncon}) 
in the case where the shell consist of an ideal fluid (\ref{eq:intSif})
we find in general

\begin{equation}\label{eq:ttgen}
K_{\tau\tau}^{\pm}=- \frac{[T_{\mu\nu}n^{\mu}n^{\nu}]}{\sigma}
-\frac{2p}{R^{2}\sigma}K_{\theta\theta}^{\pm}\pm4\pi \left(\frac{1}{2}\sigma
+2p\right),
\end{equation}
where the superscripts $\pm$ correspond to the $\pm$ in the equation, respectively. ($\sigma$ is the energy density of the surface). For a comoving observer
$K_{\tau\tau}^{\pm}$ gives the proper (normal) acceleration, i.e. 
$K_{\tau\tau}^{\pm}=a_{\mu}^{\pm}n^{\mu}_{\pm}$. Hence, the terms on the right
in eq. (\ref{eq:ttgen}) represent the forces on the shell making it deviate
from geodesic motion in $\mathcal M^{+}$ and $\mathcal M^{-}$, respectively.  
The first term is the difference in the normal force exerted on the shell 
from $\mathcal M^{\pm}$, the second term is due to the tangential pressure on the surface while the last represents the self gravity of the shell.  

We shall consider eq. (\ref{eq:ttgen}) for a flat Universe with dust and vacuum energy, i.e. $p_{cf}=0$ and $\mathcal E =1$; also we put $\Lambda_{\pm}=\Lambda$. Differentiating eq. (\ref{eq:sqeqm}) with respect to $\tau$, then using
(\ref{eq:dottau}), (\ref{eq:Kang}), (\ref{eq:geneqm}), (\ref{eq:dotmu}), 
(\ref{eq:dotMdmdr}) and (\ref{eq:tnndotM})
we find

\begin{equation}\label{eq:Rdobldot}
K_{\tau\tau}^{\pm}
=
-\zeta_{\pm}\frac{2p}{\sigma R}\sqrt{A_{\pm}+\dot{R}^{2}}
-\frac{1}{4\pi R^{2}\sigma}\left[\frac{\dot{M}_{\pm}^{2}}{\frac{\partial M_{\pm}}
{\partial R}}\right]\pm2\pi\sigma\pm8\pi p,
\end{equation}
where
\begin{equation}
K_{\tau\tau}^{\pm}=-\zeta_{\pm}
\frac{\ddot{R}+\frac{m_{\pm}}{R^{2}}+ \frac{M_{\pm}}{R^{2}}
-\frac{\Lambda_{\pm}}{3}R
+\frac{\dot{M}_{\pm}^{2}}{R\frac{\partial M_{\pm}}{\partial R}}}
{\sqrt{1-\frac{2(m_{\pm}+M_{\pm})}{R}-\frac{\Lambda_{\pm}}{3}R^{2}
+\dot{R}^{2}}}.
\end{equation}
For a test shell, $K_{\tau\tau}=0$, we recover the Friedmann eq.  
(\ref{eq:accgenR}).

\section{Conclusion}
In this paper we have discussed the embedding of a Schwarzschild mass into a cosmological model using ``curvature'' coordinates. Extending  the work by 
Gautreau \cite{gau2} we have found approximate solutions to eq. (\ref{eq:diffMpwrho})
giving the mass function M for the Universe explicitly, and we have
solved for the radial geodesics outside the embedded mass.
We have considered spacetime close to $m$, where our solutions go towards the
Schwarzschild spacetime. And far from $m$,  our solutions approach the 
FLRW Universe models. In particular we have presented
solutions for a flat Universe with vanishing cosmological constant and an equation of state $p=\omega\rho$, $\omega=constant$.
Without the embedded mass we have given solutions with and without cosmological constant for the  equation of state $p=\omega\rho$.

Using the Gautreau metric we have generalized Israel's formalism to singular shells in a Sch/FLRW background. For an arbitrary  equation of state 
for the surface the equations governing its motion are
(\ref{eq:Rdot}), (\ref{eq:dotmu}) and (\ref{eq:Rdobldot}). Equations
(\ref{eq:Rdot}) and (\ref{eq:dotmu}) are given for the general case, while 
we have only considered the acceleration (\ref{eq:Rdobldot}) for a 
flat pressurefree Universe model. An
interesting further development would be to
extend this to three-dimensional branes in a five-dimensional spacetime. 

An application of the results obtained in this paper is to study the evolution of cosmological voids, see \cite{samasa, masa, stor} and references therein. 
In \cite{stor} the collapse of a positive perturbation  leading to the Einstein-Straus model is discussed, i.e. the expansion of the voids are comoving.
In \cite{samasa,masa} a less dense region is considered. This less dense region will expand faster than the outer region  and numerical simulations show that a thin shell is formed. It would be interesting to investigate if this problem
can be analyzed analytically. Also, it would be interesting to apply our formalism to slowly rotating shells, generalizing the traetments in \cite{klein,dbd} to shells that need not be comoving with the cosmic fluid.

\end{document}